\documentclass[RNAAS]{aastex62}

\newcommand\aastex{AAS\TeX}
\newcommand\latex{La\TeX}

\graphicspath{{./}{figures/}}

\received{January 1, 2018}
\revised{January 7, 2018}
\accepted{\today}
\submitjournal{RNAAS}

\shorttitle{GBS v2.1}
\shortauthors{Jofr\'e et al.}

\newcommand{\feh}{[Fe/H]}

\begin{document}

\title{The {\it Gaia} FGK benchmark stars version 2.1}\thanks{Data available at CDS as catalog III/281}

\correspondingauthor{Paula Jofr\'e}
\email{paula.jofre@mail.udp.cl}

\author{Paula Jofr\'e}
\affil{N\'ucleo de Astronom\'ia, Facultad de Ingenier\'ia y Ciencias, Universidad Diego Portales, Ej\'ercito 441, Santiago de Chile}

\author{Ulrike Heiter}
\affiliation{Department of Physics and Astronomy,  Uppsala University,
Box 516, 75120 Uppsala, Sweden}

\author{Marcelo Tucci Maia}
\affil{N\'ucleo de Astronom\'ia, Facultad de Ingenier\'ia y Ciencias, Universidad Diego Portales, Ej\'ercito 441, Santiago de Chile}

\author{Caroline Soubiran}
\affiliation{Laboratoire d'Astrophysique de Bordeaux, Univ. Bordeaux, 
CNRS,UMR 5804, F-33615, Pessac, France}

\author{C. Clare Worley}
\affiliation{Institute of Astronomy, Madingley Road, CB3 0HA, Cambridge, UK}

\author{Keith Hawkins}
\affiliation{Department of Astronomy, The University of Texas at Austin, 2515 Speedway Boulevard, Austin, TX 78712, USA}

\author{Sergi Blanco-Cuaresma}
\affiliation{Harvard-Smithsonian Center for Astrophysics, 60 Garden Street, Cambridge, MA 02138, USA}

\author{Carlos Rodrigo}
\affiliation{Centro de Astrobiolog\'ia (CSIC-INTA), Camino Bajo del Castillo s/n, E-28692 Villanueva de la Ca\~nada, Madrid, Spain}
\affiliation{Spanish Virtual Observatory}



\section{Introduction} \label{sec:intro}

Studies of the Milky Way are experiencing a revolution thanks to the large amount of high resolution spectral data of individual stars becoming available. This revolution has just started, with on-going spectroscopic surveys delivering thousands of spectra and future surveys planned to collect millions of them. 

In order to characterise large samples of survey spectra, automatic and efficient pipelines are in constant development to carry out tasks such as measuring  equivalent widths or fitting models to data to pure data-driven approaches. In addition, the spectra are taken by different instruments, with surveys targeting different stellar populations. Hence, they are of different nature regarding resolution, signal-to-noise ratio and wavelength coverage.  

In spite of the variety of methods, the quality of the parameters obtained may be assessed in a similar way. They need to be tested against a set of reference stars, for which the parameters are known with high confidence and independently of the methodology and data employed in that particular analysis. Ideally, every survey should use the same set of reference stars, or at least have a large overlap of stars for which they can compare and scale their results. 

The GBS consists of a small but carefully selected set of FGK stars, with a large range in metallicity. A distinguishing feature of the sample is that for most of them the angular diameter has been measured with interferometry, which allows one to determine the effective temperature from the Stephan-Boltzmann relation provided the parallax and bolometric flux are known. Furthermore, a number of them are in wide binary systems or have asteroseismic measurements, thus helping us to constrain the mass and hence the surface gravity from dynamical or seismic relations.

The GBS provide the means to calibrate spectral analysis, such that the values resulting from the simplistic assumptions in spectral line formation agree with the fundamental values. The GBS have proven to be very useful to the community, with several spectral catalogues validating and calibrating their pipelines with the GBS stellar parameters. Examples are the Gaia-ESO Survey \citep{2017A&A...598A...5P}, GALAH \citep{2018MNRAS.tmp.1218B}, AMBRE \citep[e.g.][]{2016A&A...591A..81W}, RAVE \citep{2017AJ....153...75K} and recently the Apsis pipeline of Gaia-DR2 \citep{2018arXiv180409374A}. 

A series of papers has been published analysing the GBS. The first sample (v1.1) is presented in \citet[Paper I]{2015A&A...582A..49H}, and a compilation of their spectra is provided in \citet[][Paper II]{2014A&A...566A..98B}. These spectra were used for the determination of metallicity in \citet[][Paper III]{2014A&A...564A.133J}. Further determination of iron-peak and $\alpha$-element abundances was presented in \citet[][Paper IV]{2015A&A...582A..81J}. With the GBS v1.1 being used by the Gaia-ESO Survey for calibration purposes it became clear that the sample was lacking stars with \feh $ \sim -1$, which we addressed in \citet[][Paper V]{2016A&A...592A..70H}. Finally, we quantified the sources of uncertainties due to methodology in \citet[][Paper VI]{2017A&A...601A..38J}. Currently we are determining new abundances of light elements. 

\section{The GBS v2.1 Sample}
In each of the above works the GBS sample has evolved. For some stars we consider the current parameters to be too uncertain to be used as calibration/validation pillars (signified in new tables by a value of $9.99$).  
Motivated by discussions with members of the IAU working group of spectral libraries\footnote{\url{https://www.iau.org/science/scientific\_bodies/working\_groups/306/}}, making this dataset accessible with updated parameters via CDS became necessary.  We publish here the latest sample of GBS which we call GBS v2.1\footnote{\url{http://vizier.u-strasbg.fr/viz-bin/VizieR?-source=III/281}}. The stars removed from GBS v1.1 (Papers I--IV), and the metal-poor stars added from Paper V gives a total of 36 stars. We call this version 2.1 because the parameters have not been revised (they are still those derived in Papers I, III and V). 

It is our ambition to provide a new set of parameters (e.g. GBS v2.2) based on {\it Gaia} parallaxes and better measurements of angular diameters that have been published in recent years \citep[e.g.][]{2018MNRAS.475L..81K, 2018MNRAS.477.4403W}. That requires a detailed spectroscopic analysis of the kind reported in our previous papers will be published elsewhere.

Our table reports the basic observational properties and the recommended stellar parameters of the GBS v2.1 with their respective uncertainties. It further reports the abundances of iron-peak and $\alpha$-elements. The latest spectra used for our analyses can be found at the VO on the spectral libraries webpage\footnote{\url{http://svo2.cab.inta-csic.es/theory/libraries/}}. 

\section{Conclusions}
Here we present a consolidated sample of GBS v2.1. While the parameter values are published in Papers I--V, several tables presented in these papers are not available via CDS, making a cross-match with other catalogues less straightforward. That limits the applicability of the GBS.  The summary table we present via CDS and explain in this note intend to make the GBS more accessible to the community, ensuring that the latest version is used and making their application more convenient thanks to the CDS functionalities.

\acknowledgments
PJ thanks Fondecyt Iniciaci\'on Grant N. 11170174. This work is result of discussions over ISSI-BJ ``Spectral Libraries of 2020".

\bibliography{references}

\end{document}